\documentclass[prl,superscriptaddress,twocolumn,showpacs,preprintnumbers,amsmath,amssymb]{revtex4}

\usepackage{graphicx}
\usepackage{dcolumn}
\usepackage{bm}

\begin{document}
\title{Wigner crystallization in  Na$_3$Cu$_2$O$_4$ and
Na$_8$Cu$_5$O$_{10}$ chain compounds
}
\author{P. Horsch, M. Sofin, M. Mayr, and M. Jansen }
\affiliation{Max-Planck-Institut f\"ur Festk\"orperforschung,
             Heisenbergstrasse 1, D-70569 Stuttgart, Germany}

\date{\today}

\begin{abstract}

We report the synthesis of novel edge-sharing chain systems
Na$_3$Cu$_2$O$_4$ and Na$_8$Cu$_5$O$_{10}$, which form insulating states 
with commensurate charge order. 
We identify these systems as one-dimensional Wigner lattices, where the 
charge order is determined by long-range Coulomb interaction and the number
of holes in the d-shell of Cu.
Our interpretation is supported by  X-ray structure data as well as 
by an analysis of magnetic susceptibility and specific heat data.
Remarkably, due to large second neighbor Cu-Cu hopping, these systems allow
for a distinction between the (classical) Wigner lattice 
and the $4 k_F$  charge-density wave of quantum mechanical origin.

\end{abstract}
\maketitle


The role of strong electron correlations and the 
concomitant appearance of {\it spatially modulated 
charge structures} constitutes a central issue in 
current solid state physics \cite{Orenstein00}. 
The most prominent example are charge stripes in high-T$_c$ superconductors, 
which have been discovered first as static modulations 
in the CuO$_2$ planes of  La$_{1.6-x}$Nd$_{0.4}$Sr$_x$CuO$_4$\cite{Tranquada95}.
The high-$T_c$ enigma
has at the same time stimulated the florishing field
of low-dimensional cuprates composed of Cu-O chains and 
ladders \cite{Dagotto96} 
with the hope to gain new insights overlooked so far in the layered cuprates.
One important aspect 
is the role of long-range Coulomb interaction in strongly correlated systems, 
which gives rise to metal/insulator stripe structures in organic charge-transfer
compounds\cite{Kumai99}.


Following a recently discovered new route in the synthesis of 
alkalioxometallates \cite{Trinschek99}, 
we were able to synthesize several members of a new class of quasi-1D cuprates
Na$_{1+x}$CuO$_2$. 
These {\it intrinsically doped} edge-sharing chain systems 
provide a unique opportunity to study
the condensation of charge order (CO) at high temperature and the formation of
spatially modulated Heisenberg spin systems at low temperature.
Edge-sharing chains are also building blocks of the intensively studied
Sr$_{14-x}$Ca$_{x}$Cu$_{24}$O$_{41}$ system; due to exchange of electrons 
with ladders, however, the degree of doping is difficult to determine \cite{Osafune97}.
Here, we argue that these doped chains can be understood as realizations of 
one-dimensional Wigner lattices (WL) \cite{Wigner34}, 
as introduced by Hubbard \cite{Hubbard78} in the late 70's
in connection with TCNQ charge transfer salts. 
He suggested that the distribution of electrons is controlled by the Coulomb 
interaction rather
than by the kinetic energy ($\sim$ band width), such that they form
a {\it generalized Wigner lattice}  on the underlying TCNQ chain structure.
This view suggests a strikingly different nature of charge excitations,
namely as domain walls with fractional charge rather than particle-hole 
excitations as in common metals and semiconductors.
Hubbard's proposal, however, can be challenged on the grounds that 
the resulting periodicity of charge modulation can alternatively be explained
by a $4k_F$ charge density wave (CDW)  \cite{Solyom79,Voit94,Gruener94}
arising from short-range interactions alone and an instability of the
Fermi surface, where $k_F$ denotes the Fermi momentum.

The edge-sharing arrangement 
of CuO$_4$ squares  meets the WL criterion of small 
band width in an optimal way due to the almost 90$^o$ Cu-O-Cu bonds (Fig.1).
Unexpected complexity is added
because, apart from a small nearest-neighbor
hopping matrix element $t_1$, there is also a second neighbor hopping $t_2$, which 
is larger as a consequence of the structure.
While this unusual feature does not affect the classical WL order imposed by the
Coulomb interaction, it changes the Fermi surface topology, and thereby 
allows to distinguish the WL from the CDW on the basis of the
modulation period.

These systems provide a first example where an unambiguous distinction between
the generalized WL and a Fermi surface related $4 k_F$ CDW is possible.
We also show that for these edge-sharing compounds even the magnetic and 
thermodynamic properties can only be explained by invoking a WL
ground state emerging from the truly long-ranged Coulomb interaction.

%

{\it Samples} were prepared by the azide/nitrate route \cite{Trinschek99}.
As a source for the alkalimetal component, mixtures of the respective 
alkali azides and nitrates (or nitrites) are used instead of the alkalioxides.
Conveniently, besides the metals' ratio, also the oxygen content and thus the degree and 
kind of doping of the desired product can be effectively controlled using the 
weighed portions of the starting materials. Following this procedure 
the title compounds Na$_3$Cu$_2$O$_4$ $(x=1/2)$ and Na$_8$Cu$_5$O$_{10}$ ($x=3/5$) 
have been prepared 
as microcrystalline, pure phases in gram-amounts. 
The new 
oxocuprates (II/III) belong to the compositional series Na$_{1+x}$CuO$_2$, with the end 
members NaCuO$_2$ \cite{Hestermann69} and the still elusive Na$_2$CuO$_2$. 
The most prominent structural feature, 
common to all representatives known thus far, is a one-dimensional polyanion 
CuO$_2^{n-}$ constituted of CuO$_4$-square units sharing edges in 
trans-position. These anionic entities are embedded by sodium ions which achieve 
coordination numbers of 4-6 with Na-O bond lengths ranging 
from 227 to 279 pm. The geometric data as determined by single crystal 
structural analyses give clear evidence for a 
CO at the Cu sites (Fig. 1). 
Based on the Cu to O bond lengths, one can unambiguously distinguish 
Cu$^{3+}$ and Cu$^{2+}$ sites. 
The way of linking the primary structural units together with the 
variations of the copper to oxygen distances inevitably leads to deviations of the 
O-Cu-O angles from the ideal 90$^o$. As monitored by differential scanning calorimetry 
measurements CO disappears above 
the WL melting temperature $T_m=455$ and 540 K 
for Na$_3$Cu$_2$O$_4$ and Na$_8$Cu$_5$O$_{10}$, respectively.
DC and AC conductivity measurements show also a clear transition from
an Arrhenius behaviour below $T_m$ to an almost temperature independent conductivity
regime above $T_m$.

\begin{figure}
\includegraphics[width=5.4cm]{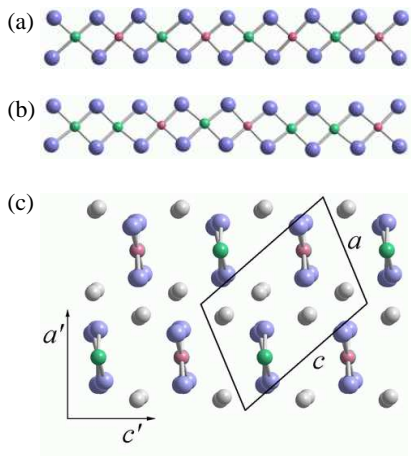}
\vspace*{0mm}
\caption{\label{fig:structure1}
Structure of edge-sharing
copper-oxygen chains along $b$-direction in  Na$_3$Cu$_2$O$_4$ (a) and  
Na$_8$Cu$_5$O$_{10}$ (b), 
where Cu$^{2+}$ and Cu$^{3+}$ are marked by green and red circles, 
and O$^{2-}$ (Na$^+$) by large blue (grey) circles, respectively. 
(c) View on the $a-c$ plane of  Na$_3$Cu$_2$O$_4$.
For convenience we introduce a cartesian system $a',b'||b,c'$ in addition
to crystallographic coordinates. 
}
\end{figure}


{\it For a theoretical analysis} one has to recognize that Cu$^{2+}$ is in a $d^9$ 
configuration
with spin 1/2, while Cu$^{3+}$ is in a $d^9$-ligand hole ($d^9 L_h$) state,
also known as Zhang-Rice singlet state \cite{Zhang88}.
In contrast to high-T$_c$ cuprates the edge-sharing geometry (Fig.1(a,b)) 
leads to strongly reduced hopping matrix elements. 
This sets the stage for the long-range Coulomb force as dominant interaction
\begin{equation}
H_{Coul}= U\sum_{i} n_{i,\uparrow} n_{i,\downarrow} 
+\sum_{i, l\geq 1} V_{l} n_i n_{i+l}  ,
\end{equation}
where the on-site interaction $U$ 
suppresses charge fluctuations involving Cu$^{1+}$ ($d^{10}$) configurations.
Here we associate the $d^9L_h$ ($d^9$,$d^{10}$) ionization state with
0 (1,2) electrons, respectively, and $n_{i,\sigma}$ ($\sigma=\uparrow,\downarrow$)
counts the number of electrons with spin $\sigma$, while 
$n_i=n_{i,\uparrow}+n_{i,\downarrow}$. 
The Coulomb interaction $V_{l}$ in general is screened by the polarization of
neighbouring chains as well as by core electrons \cite{Hubbard78}. 
Here we assume 
a generic Coulomb law  $V_l=\frac{V}{l}$, $l=1,2,....$\cite{periodic}. 
Crucial for the following is that the interaction is long ranged and
convex, i.e., $V''_l=V_{l-1}-2 V_l+V_{l+1}$$>$$0$.

\begin{figure}
\includegraphics[width=6.0cm]{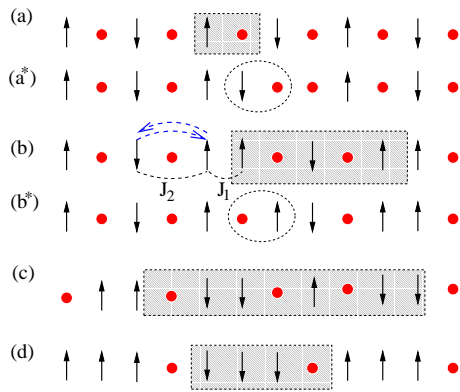}
\caption{\label{fig:wigner1}(color online)
Wigner charge order resulting from Coulomb repulsion
and associated modulated Heisenberg spin structure
for $x=1/2, 3/5, 5/8$ and $3/4$ doping (a-d). 
The spin-1/2  of Cu$^{2+}$ (arrows) is responsible 
for magnetism; Cu$^{3+}$ (circles) is nonmagnetic.
The spin arrangement is that expected for ferromagnetic exchange $J_1$
and antiferromagnetic $J_2$, 
where the two excitations $d^9\cdot d^9 \leftrightarrow d^9L_h\cdot d^{10}$
contributing to $J_2$ are indicated by dashed arrows in (b).
The charge unit cells (shaded) 
contain 2, 5, 8, 4 sites, respectively.
The structures (a) and (b) are realized in   Na$_3$Cu$_2$O$_4$  and
Na$_8$Cu$_5$O$_{10}$, respectively, and the dashed circles in (a$^*$) and (b$^*$)
indicate charge excitations $\propto V$ in these structures. 
(d) Typical modulation for charge stripes in cuprates at doping 1/8.
 }
\end{figure}

For commensurate doping concentration $x=m/n$ the interaction $V_l$
selects a particular CO pattern\cite{Hubbard78}. 
This pattern is
immediately obvious for filling fractions $x=1/2$ and 3/4 (Fig. 2(a,d))
which involve an equidistant arrangement of the Cu$^{3+}$ sites
(red circles in Fig. 2). 
For a general ratio $x=m/n$ this leads to complex structures with unit cell size 
$n$ (in units of the Cu-Cu distance $b'=1$).
In case of $x=3/5$ we encounter in Fig. 2(b)  the charge order observed for  
Na$_8$Cu$_5$O$_{10}$.
Charge localization, however, is not perfect in Wigner insulators 
as electrons still undergo 
virtual transitions to neighboring sites (Fig. 2(a$^*$,b$^*$))
in order to retain partially their {\it kinetic energy}. 
The energy of the lowest excitations and the impact of kinetic energy
depend strongly on $x$, e.g., the energy
of the excitation in Fig. 2(a$^*$) relative to the ground state Fig. 2(a) is 
$\sim V''_2$
while the excitation for $x=3/5$ in Fig. 2(b$^*$) is $\sim V''_5$, about an order of 
magnitude smaller. 
To investigate the role of kinetic energy
we explore the dynamics of electrons starting from the 
1$D$ {\it Hubbard-Wigner model} $H_{HW}=H_{Coul}+H_{Kin}$ \cite{Hubbard78},
where
\begin{equation}
H_{Kin}=-\sum_{i,l,\sigma} t_l(c^+_{i+l,\sigma}c_{i,\sigma}
+c^+_{i,\sigma}c_{i+l,\sigma})
\end{equation}
describes the hopping of an electron with spin $\sigma$.  
Due to the almost 90$^o$ Cu-O-Cu angle the hopping  $t_1$ 
between nearest neighbor Cu sites results mainly 
from direct $d-d$ exchange, while $t_2$ originates from
hopping via a Cu-O-O-Cu path \cite{Mizuno98} (Fig. 1), leading to the remarkable
fact $|t_2|>|t_1|$.
We adopt here as typical values $t_1\sim 63$ meV, $t_2\sim 94$ meV,
derived from ab-initio  band structure calculations
for the Cu$^{2+}$ edge-sharing reference system Li$_2$CuO$_2$ \cite{Weht98}.  
These values are indeed much smaller than our estimates for $U\sim 3.8$ eV and 
$V\sim 1.5$ eV based on optical data for Li$_2$CuO$_2$ \cite{Mizuno98}.

\begin{figure}
\includegraphics[angle=-0,width=8.0cm]{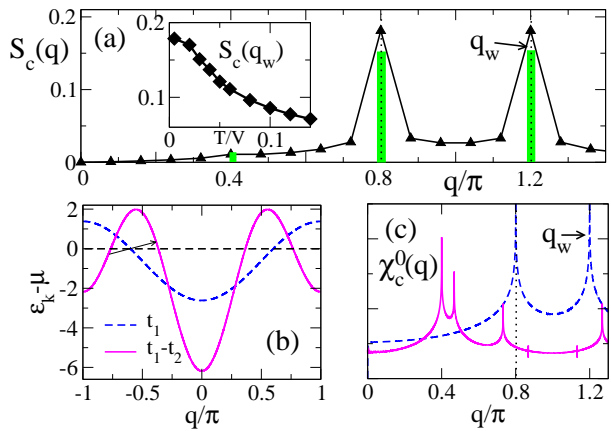}
\caption{\label{fig:chiofq}(color online)
(a) Static charge structure factor  $S_c(q)$ calculated for
interacting spinless fermions and  $t_1=0.04 V$,  $t_2=0.06 V$ at $x=0.6$
indicates the instability towards WL ordering with modulation $q_w$=1.2$\pi$
(solid line as guide to the eye). 
Inset shows the temperature dependence of  $S_c(q_w)$.  
(b) Comparison of electron dispersions for a system
with nearest neighbor hopping $t_1(=1)$ only (dashed line)
and for $t_1=1$, $t_2=1.5$ (solid line) with  
the Fermi energy  indicated by a horizontal line. 
(c) While the charge susceptibility $\chi_c^0(q)$ for noninteracting
spinless fermions in the $t_1$ model shows 
a logarithmic divergence at $2 k_{SF}=q_w$, the singularities of the $t_1$-$t_2$ 
model are at different momenta.
}
\end{figure}
Using exact diagonalization 
we have calculated the static structure factor $S_c(q)$
for chains up to 25 sites. 
The peaks of $S_c(q)$ 
at $q_w=1.2\pi$ (and at $0.8\pi$=$2\pi$-$q_w$) are 
characteristic for a WL modulation at $x=3/5$ (Fig. 3(a)). 
 Since the magnetic energy scale is much smaller than the Coulomb interaction $V$, 
the charge structure can be determined by disregarding spin degrees of freedom, 
namely in terms of spinless fermions (SF).
The SF-CDW arises as an instability 
due to low energy scattering between the two Fermi points at $\pm k_{SF}$ which 
lead to a singularity in the charge susceptibility $\chi_c^0(q)$ at $q=2 k_{SF}$(=4$k_F$). 
Subsequent inclusion of interactions is expected to change the character of a
singularity, yet not the momentum at which the singularity occurs.
This explains the origin of the $4 k_F$ CDW instability 
in interacting 1D systems and, as $4 k_{F}=2\pi m/n$,
the coincidence with the modulation period $n$ of the WL.

The equivalence of WL and CDW periodicity disappears when $t_2$$>$$\alpha$$t_1$ 
($\alpha=0.38$ at $x=0.6$); then there are
four instead of two Fermi points leading to new singularities in 
$\chi^0_c(q)$ (Fig.3(b,c)) and to a shift of the original singularity away from 
$4 k_F=q_w$.
In this case the standard tools of many body theory 
such as  bosonization \cite{Voit94} suggest a change of the modulation period.
This, however, is not reflected in the experiment and is also not
observed in our numerical results for $S_c(q)$ (Fig. 3(a)).
In fact, the singularity remains at $q_w$ up to a value $t_2\sim 4 t_1$,
a striking manifestation that the structure is robust and controlled  
by the long range Coulomb interaction \cite{note}.    
The calculated $S_c(q)$ shows besides the main peak at $q_w$ a weak
higher harmonic feature at $q=2 q_w-2\pi$, consistent
with the relative intensities of structural reflexes calculated from the experimentally
determined Cu ion positions (vertical bars in Fig. 3(a)).
Further results (inset Fig.3(a)) reveal that WL correlations persist up
to temperatures $k_B T\sim 0.05 V$, consistent with experiment.


{\it The magnetic susceptibility} $\chi(T)$ was measured in the temperature range
from 5 K to 350 K using a SQUID magnetometer. 
The data  (Fig. 4(a))
reveal strikingly different temperature dependences for the two
compounds: $\chi(T)$ for  Na$_3$Cu$_2$O$_4$ displays some similarity with
a nearest neighbor Heisenberg antiferromagnetic chain \cite{Johnston00}, whereas 
the Na$_8$Cu$_5$O$_{10}$
data show continuous increase down to low temperature until its
maximum near 25 K is reached. Both systems reveal antiferromagnetic correlations, 
yet magnetic order is observed 
only for Na$_8$Cu$_5$O$_{10}$ at $T_N = 23.5$ K. For a calculation of $\chi(T)$
we assume that the spins remain fixed at their positions ${\bf R}_i$ as given by the
structural analysis and by $H_{HW}$ (Fig.2). 
This leads to a generalized Heisenberg model
\begin{equation}
H_{Heis}=\frac{1}{2}\sum_{i,j} J({\bf R}_i-{\bf R}_j)
{\bf S}_i\cdot {\bf S}_j,
\end{equation}
where the exchange constants depend on the distance $|{\bf R}_i-{\bf R}_j|$
between the spins and
on the direction (parallel to the chains or perpendicular). 
For $x=0.5$ 
only the exchange constants $J_2, J_4, ....$ along the chains contribute, 
while for $x=0.6$ $J_1, J_2, J_3, ....$ are relevant. 
Apart from the modulated spin pattern, 
superexchange \cite{Anderson59} in WL's
shows further novel features, namely  fluctuations (a) of spin positions
and (b) of antiferromagnetic exchange integrals $J_l\sim 4 t_l^2/(U+\Delta(V))$ 
due to the
low energy charge fluctuations. The energy shifts $\Delta(V)$ due to
$V_l$ depend on the WL structure, and are in general different
for left (right) scattering processes (see Fig. 2(b)).
The exchange constants may be estimated from the parameters specified above.
The largest coupling is $J_2\approx 100-200$ K resulting
from superexchange via a Cu-O-O-Cu path.
Because of the almost 90$^0$ Cu-O-Cu bond angle (13)
$J_1$ exchange is smaller. 
Yet in the case of $J_1$ there are additional exchange contributions, which have 
to be taken into account. The most important of these involve  O $p^4$ 
configurations with  
Hund interaction on oxygen \cite{tornow98}, such that the total $J_1$ 
may become negative.

\begin{figure}
\includegraphics[width=6.9cm]{horschFIG_4a.eps}
\includegraphics[width=6.9cm]{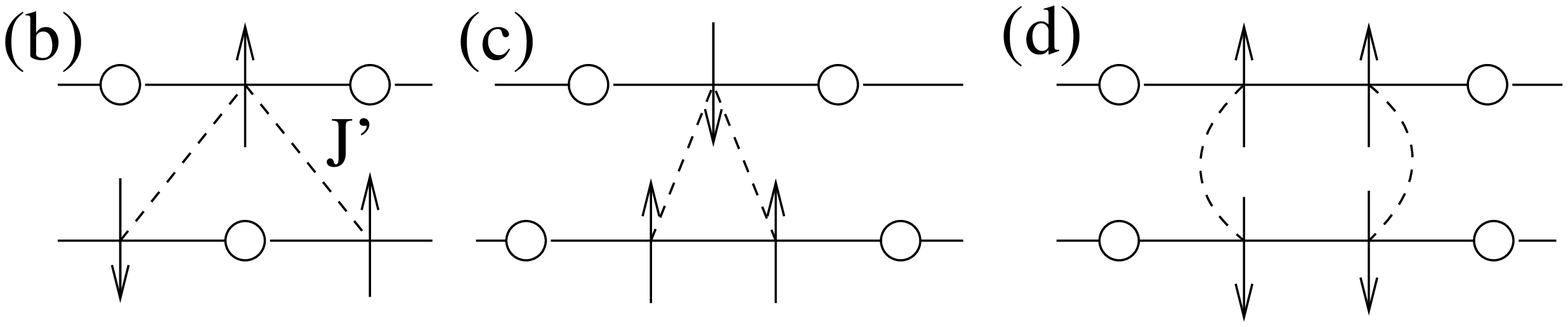}
\caption{\label{fig:chi}(color online)
(a) Temperature dependence of susceptibility $\chi(T)$ 
for Na$_3$Cu$_2$O$_4$ and $\chi/3$ for Na$_8$Cu$_5$O$_{10}$.
Multiplication by $1/3$ in the latter case  provides a comparison per spin. 
Theoretical results calculated by FTD are indicated by lines (exchange constants
are given in the text).
Sketch of interchain interactions used in the calculation of  $\chi(T)$: 
For  Na$_3$Cu$_2$O$_4$ (b) the magnetic exchange $J'$
is frustrated due to triangular coordination, while it is not frustrated 
for  Na$_8$Cu$_5$O$_{10}$ both along $a'$-direction (c) and $c'$-direction (d).
}
\end{figure}

We have used finite temperature diagonalization (FTD) \cite{Jaklic00} to 
calculate $\chi(T)$ 
for  Na$_3$Cu$_2$O$_4$ studying chains up to $N=48$ sites (24 spins). 
Good agreement with experiment was achieved with  $J_2=172$, $J_4=17$ K 
and 1.97 for the g-factor. 
The small antiferromagnetic interchain coupling $J'=34$ K (Fig. 4(b))
is frustrating as the correlations on  $b'$ chains are  antiferromagnetic.
The interaction $J'$ 
has been taken into account in a mean-field approach\cite{Johnston00}
$\chi(T)=\chi_s(T)/(1 + z' |J'| \chi_s(T))$,
where $\chi_s(T)$ is the result for a single chain, while $z'$
counts the number of spins on neighboring chains ($z'=4$). 
The frustrated geometry implies a spin gap $\Delta_s$ \cite{White96}.
For the present parameter set we obtain  $\Delta_s\sim 4$ K,
consistent with our analysis of the specific heat data.  
A thorough analysis of the Curie contribution, 
using specific heat and susceptibility data independently, 
reveals that there are about $\sim$1$\%$ impurities per Cu ion present in 
both Na$_3$Cu$_2$O$_4$ and Na$_8$Cu$_5$O$_{10}$ \cite{Mayr_p}.

In the case of Na$_8$Cu$_5$O$_{10}$ the correlations along the $b$-chains
turn out to be of {\it ferrimagnetic} nature.
A full FTD calculation of a double chain was necessary to
account for the AF interchain exchange $J'$ along $c'$-direction 
(Fig. 4(d)), while the coupling of
chains along $a'$-direction (Fig. 4(c)) 
was treated within mean-field. 
This implies ferrimagnetic $a'$-$b'$ planes, which are antiferromagnetically coupled
along $c'$-direction.
A satisfactory description was found for $J_1=-22$ K, $J_2=113$ K, $J'=21$ K 
and $g=2.0$.  
The ferromagnetic interaction $J_1$ is the origin of the strong enhancement of 
$\chi(T)$ in the $x=0.6$ system. 
The reduced value of $J_2$ 
can be explained by low-energy charge
fluctuations (Fig. 2(b$^*$)), which involve a fluctuation of spin positions,
i.e., $J_2^{eff}\simeq (1-p)J_2+p J_1$ and a larger probability $p$ than for $x=1/2$.

The strong variation of  $\chi(T)$ in Na$_8$Cu$_5$O$_{10}$ 
should be interpreted as a crossover between two regimes. While at high temperature
$\chi(T)$ is controlled by 
the individual spins,  at low temperatures only {\it effective} spins 1/2, 
formed
by three spins which are  coupled by the largest interaction $J_2$,
are relevant  (Fig. 2(b)).
The composite nature of the effective spins and their internal energy level 
structure becomes relevant
in the intermediate temperature region.
The interaction between these effective spins is renormalized to $J_1^{eff}=\frac{4}{9}[J_1-J_3+\frac{1}{4}J_5]$. Thus,
the peculiar low-T behavior of $\chi(T)$ is controlled by an interplay of 
the small $J_1^{eff}$ between effective spins, resulting from WL order,
and the interchain interactions.


Finally, the new compounds provide a unique opportunity to study the competition 
between two entirely different states, the classical WL dictated by the long-range 
Coulomb interaction and the CDW of quantum mechanical origin, i.e., resulting
from a Fermi surface instability.
These materials highlight  
the importance of long-range Coulomb interaction in strongly correlated systems, 
--- and provide a one-dimensional test ground for the study of
charge stripe formation.

We thank E. Br\"ucher and G. Siegle for the susceptibility and 
specific heat measurements
and L. Capogna, B. Keimer, W. Metzner and A.M. Ole\'s for discussions.


\begin{thebibliography}{99}
\bibitem{periodic}
In the case of periodic boundary conditions we define
$V_l=max[V/l,V/(N-l)]$, where $1<l<N$.
\bibitem{note}
For small hopping $S_c(q)$ is peaked at $q_w$ irrespective of the signs of $t_1$ and $t_2$!
\end{thebibliography}
\end{document}